\documentclass[10pt]{JHEP3}

\usepackage{amssymb,amsmath}
\usepackage{graphics}
\usepackage{epsfig}
\usepackage{amsfonts}

\usepackage{amscd}

\def\be{\begin{eqnarray}}
\def\ee{\end{eqnarray}}

\def\e{\epsilon}

\def\half{\frac{1}{2}}
\def\d{\partial}
\def\a{\alpha}

\newcommand{\idn}{{1\relax{\kern-.35em}1}}

\def\a{\alpha}

\def\e{\epsilon}

\preprint{ WIS/11/07-JUL-DPP }

\title{Planar Quark Scattering at Strong Coupling and Universality}
\author{Zohar Komargodski$^1$, and Shlomo S. Razamat$^2$\\
$^1$ Department of Particle Physics,
The Weizmann Institute of Science,
Rehovot 76100, Israel\\
$^2$ Department of Physics,
Technion, Israel Institute of Technology,
Haifa 32000, Israel\\
\email{Zkomargo@weizmann.ac.il},
\email{razamat@physics.technion.ac.il} }

\abstract{We discuss scattering of fundamental matter in the
planar and strong coupling limit via the $AdS/CFT$ correspondence,
generalizing the recently proposed calculation for adjoint matter
due to Alday and Maldacena \cite{Alday:2007hr}. Color decomposition
of quark amplitudes is a key property allowing to repeat the
procedure in the case of fundamental matter and to derive the
relation of these strong coupling amplitudes to minimal area problems. We present the results for two different $D3-D7$
systems, one is only conformal in the planar limit and the other
is exactly conformal. Our results suggest a universal behavior of
scattering amplitudes at strong coupling and planar limit (both
for gluons and quarks).}


\begin{document}

\section{Introduction and summary}

In the last decade great amount of research was devoted to the study of the
$AdS/CFT$ correspondence \cite{ADS}. This
correspondence provides us, among other things, with a technique
to calculate correlators of composite gauge invariant operators in
strongly coupled quantum field theories. On one side of the
correspondence we have a gauge theory and on the other side a
closed string theory, which is usually considered in the
supergravity approximation. A priori any knowledge of correlators
of the basic gauge fields is lost on the string theory side of the
correspondence, as these are associated with open string degrees
of freedom absent in a closed string theory.

Recently, a very interesting proposal has been put forward
by Alday and Maldacena \cite{Alday:2007hr} to recover the
information about correlators of the elementary gauge fields from
the gravity dual.\footnote{See \cite{AM_F} for more developments.} The main input into the success of this program
is that these correlators of gluons can be decomposed as products
of tensors in the color space, and gauge invariant parts (in the
sense that null states decouple) referred to as the "reduced
amplitudes". This fact goes under the name of color decomposition.

Based on this, Alday and Maldacena have suggested a concrete way to compute gluon scattering amplitudes of
strongly coupled planar ${\mathcal N}=4$ SYM theory. In this limit, the scattering amplitudes are shown to be
equivalent to computations of areas of surfaces whose boundary consists of light-like segments. The striking
success of this prescription is the agreement of the result \footnote{In the case of $4$-pt scattering,
\cite{Alday:2007hr} have used a conformal transformation of a previously known solution due to
\cite{Kruczenski:2002fb} in order to obtain an explicit minimal surface.}
 with a conjectured form of these
amplitudes due to Bern, Dixon and Smirnov \cite{Bern:2005iz} (see also \cite{Anastasiou:2003kj}). Thus, it is a
strong evidence in favor of this conjecture holding for all values of the coupling.

In this note we discuss the planar contribution to the strong coupling quark scattering amplitudes in two different
$\mathcal{N}=2$ supersymmetric gauge field theories. The first is
$\mathcal{N}=4$ $U(N)$ gauge theory deformed by adding an
$\mathcal{N}=2$ hypermultiplet in the fundamental representation
\cite{Karch:2002sh}. The second is a conformal $\mathcal{N}=2$
theory with a symplectic gauge group \cite{Aharony:1996en}.

The results of \cite{Alday:2007hr} suggest universality of gluon
scattering in a certain class of strongly coupled large $N$ conformal field theories.
This class consists of conformal field theories dual to a semiclassical string
theory background of the form $AdS_5\times W$. In this case, it is quite straightforward to repeat the
procedure of \cite{Alday:2007hr}, obtaining the same result.\footnote{In \cite{Alday:2007hr}, the
classical solution does not involve the $S^5$ as it can only
increase the action. This is why one expects the result to be
independent of the transverse space.} It is unlikely that all
these conformal theories have the same planar scattering
amplitudes perturbatively in the 't-Hooft coupling (if this limit exists);\footnote{One
has to keep in mind that there are possibly many planar
equivalences (for a review see \cite{Armoni:2004uu}). } the
universality occurs in strong coupling.

It is then natural to inquire whether similar universality holds for other matter fields which may appear in
such theories (either exactly conformal theories or conformal only for large $N$). We analyze the case of
fundamental representations in this note, and again find the same results for a-priori different strongly
coupled planar theories. It is intriguing to understand from the field theory point of view why this
universality occurs.

We wish to emphasize that neither gluon nor quark scattering is a
well defined observable due to the IR divergences. However, if one
takes into account gluon emission from the external lines then
both become IR safe. It is, nevertheless, interesting to compute
the strict $n$ point function as it is a building block in many IR
safe computations.

This note is organized as follows. In section \ref{sect_codec} we
describe the way color decomposition works in the case of
fundamental matter. In section \ref{sect_quarks} we review how one
can add quarks to ${\mathcal N}=4$ theory. In section
\ref{sect_Tdual} we repeat the necessary steps for calculating the
reduced amplitude of quarks scattering in the model of section
\ref{sect_quarks}. In section \ref{N=2conf} we discuss another
model with "quark" fields and repeat the argument once again.

\section{Color decomposition of quark scattering}\label{sect_codec}
Let us briefly review color decomposition of gluon amplitudes
\cite{Berends:1987cv,Mangano:1987xk} (see \cite{Mangano:1990by}
for a review). The general amplitude depends on the color indices,
momenta and helicities of each in-going gluon (we assume all of
them are in-going for simplicity). These quantum numbers are
denoted by $a_i$,$k_i$,$h_i$ respectively. It can be shown that
planar amplitudes factorize as
\begin{gather}\label{gluon factorization}
   \mathcal{M}^{gluons}_n(k_1,h_1,a_1;k_2,h_2,a_2;...;k_n,h_n,a_n)=\cr=\sum_{\sigma\in S_n/\mathbb{Z}_n}Tr(T^{a_{\sigma(1)}}T^{a_{\sigma(2)}}...T^{a_{\sigma(n)}})A^{gluons}_n(k_1,h_1;...;k_n,h_n),\end{gather}
where the sum is over permutations which do not differ by cyclic permutations.
In this paper, we consider $U(N_c)$ gauge theory with minimally
coupled quarks in the fundamental representation and anti-quarks
in the anti-fundamental representation (there may also be
couplings to adjoint matter). Hence, one should first generalize
(\ref{gluon factorization}) to this case
\cite{Mangano:1988kk,Kosower:1988kh}.\footnote{A more general
problem includes quarks, anti-quarks and gluons as external
states. The color structure in this case can be successfully
analyzed as well - for a review see \cite{Mangano:1990by} and
references therein.}

\begin{figure}[htbp]
 \begin{center}
 $\begin{array}{c@{\hspace{0.6in}}c}
\epsfig{file=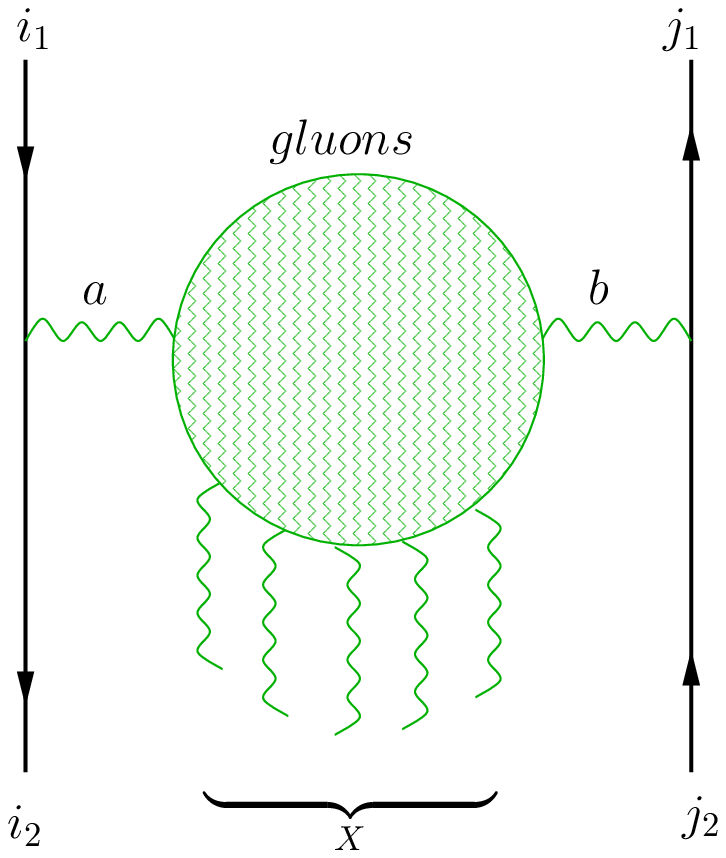,scale=0.5} & \epsfig{file=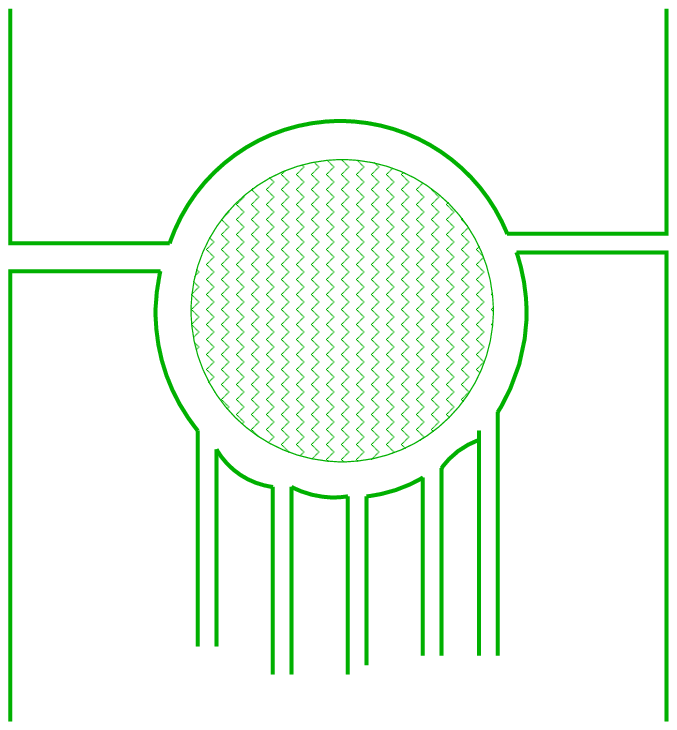,scale=0.5}
\\ [0.2cm]
(a)&(b)\\ [0.2cm]
\end{array}$
\caption{Example of quark scattering diagram, (a) the Feynman diagram (b) its double line notation.} \label{quarks1}
\end{center}
\end{figure}

For simplicity, we assume that there is only one flavor (this
assumption can be easily generalized). For the combinatorial
analysis, we also assume that all the momenta are ingoing. Thus,
we actually describe scattering of $n$ quarks and $n$ anti-quarks.
One would like to calculate the connected amplitude for such a
scattering process
              \begin{equation}\label{amplitude}\mathcal{M}^{quarks}_n(k_1,\epsilon_1,a_1 ; k_2,\epsilon_2,a_2;...;k_n,\epsilon_n,a_n|r_1,\bar\epsilon_1,\bar a_1 ; r_2,\bar\epsilon_2, \bar a_2;...;r_n,\bar\epsilon_n, \bar a_n   ),\end{equation}
where $k_i$, $r_i$ are the momenta of the quarks and the anti-quarks
respectively, $\epsilon$,$\bar\epsilon$ are their respective
polarizations and $a$, $\bar a$ are their respective color
 indices.  This is, of course, a formidably hard problem. We would
like to show that there is a very convenient way of taking into
account the color dependence of planar diagrams, analogous to
(\ref{gluon factorization}). The first
simplification due to the planar limit is that  quark
loops are sub-leading in this approximation. Hence, if a propagator in a Feynman diagram is not a ``continuation'' of an external quark
propagator it must be an adjoint field. Let us abbreviate (\ref{amplitude})
as
\begin{equation}\label{abbreviated amplitude}
    \mathcal{M}^{quarks}_n(1;2;..;n | \bar 1;\bar 2;...;\bar n).
\end{equation}
To understand the color flow it is best to think in 't Hooft's double line notation, but let us do an example explicitly. Consider the diagram depicted in figure \ref{quarks1}(a). The indices are color indices, and the arrows are the color flow direction consistent with planarity.
The color structure of this amplitude is given by
\begin{equation}
    \sum_{a,b}T^a_{i_1i_2}\mathcal{M}^{adjoint}_n(k_1,h_1,a;k_2,h_2,b;...)T^b_{j_2j_1}.
\end{equation}
Adjoint amplitudes color decomposition simplifies the color dependence of the expression above to
\begin{equation}
    \sum_{a,b,X}T^a_{i_1i_2}Tr(T^aT^bX)T^b_{j_2j_1},
\end{equation}
where X encompasses all the color information associated to the emitted gluons other than $a$ and $b$.
Finally, we can use some simple $U(N_c)$ identities in a convenient normalization
\begin{gather}\label{identities}
Tr(T^aT^b)=\delta^{ab},\cr
\sum_{a=1}^{N_c^2}T^a_{ij}T^a_{kl}=\delta_{il}\delta_{jk},
\end{gather}
where $T^a$'s are $n\times n$ hermitian matrices to get
\begin{equation}
    \sum_{a,b,X}T^a_{i_1i_2}Tr(T^aT^bX)T^b_{j_2j_1}=\delta_{i_1j_1}\sum_X(X)_{i_2j_2}.
\end{equation}
The important term is $\delta_{i_1j_1}$. Drawing the diagram in double line notation, as in figure \ref{quarks1}(b), this delta function is obvious since there is a line along which the color $i_1$ flows, and eventually connects to $j_1$. Similarly, for each Feynman diagram one should follow the flow of the external color. This pairs the quarks and the anti-quarks in a unique way and gives a delta function for each such pair. Consequently, the natural color decomposition of (\ref{abbreviated amplitude}) is
\begin{gather}\label{decomposed amplitude}
    \mathcal{M}^{quarks}_n(1;2;..;n | \bar 1;\bar 2;...;\bar n)=\cr=\sum_{\sigma\in S_n} \delta_{a_1 \bar a_{\sigma(1)}}\cdot\cdot\cdot\delta_{a_n \bar a_{\sigma(n)}} A^{quarks}_n(1;..;n |\overline{\sigma(1)};...;\overline{\sigma(n)}).
\end{gather}
The indices of the function $A^{quarks}_{n}$ stand for momenta and
polarization quantum numbers only (it is color independent by
construction). Two remarks are in order. First, the result
(\ref{decomposed amplitude}) is a very general one. A slight
generalization of the construction above implies that the set of
tensors spanned by
\begin{equation}\label{tensors}\delta_{a_1 \bar
a_{\sigma(1)}}\cdot\cdot\cdot\delta_{a_n \bar
a_{\sigma(n)}}\end{equation} is a suitable set of tensors to
expand any amplitude of quark scattering, even beyond the planar
limit.\footnote{In the case of gluon scattering, the set of
tensors spanned by the single traces is not sufficient beyond the
planar limit. One needs to include multi-traces as well.} Intuitively, the reason is that
following external fundamental color line in the double line
notation, even for a diagram of some non zero genus, eventually
leads one to an external anti-quark producing the required delta
function. However, this does not mean that the expansion
(\ref{decomposed amplitude}) is {\it useful} beyond the planar
limit. The point is that the reduced amplitude, $A_n$, is not
guaranteed to be gauge invariant in general (in the sense of null states decoupling). In the planar limit,
one can prove that reduced amplitude are gauge invariant by
utilizing the approximate orthogonality of the tensors
(\ref{tensors}) (for details see \cite{Mangano:1990by}). This
guarantees their linear independence in leading order.

\section{Adding fundamental matter to ${\mathcal N}=4$} \label{sect_quarks}

  We wish to add fundamental matter to ${\mathcal N}=4$ so that we can study its scattering amplitudes. We briefly review here two ways in which
  quarks can be added to ${\mathcal N}=4$ in the framework of
  $AdS/CFT$. Another way to do it is explained in section
  \ref{N=2conf}.

   The simplest way to add quarks is by
  going to the Higgs branch and breaking $SU(N+M)$ to $SU(N)\times SU(M)$. We take $N$ large but $M$ finite, and take the near horizon limit of the $N$
  branes considering the remaining $M$ branes  in the probe approximation. No SUSY is broken and the fundamental matter comes from strings stretching
  between the probe D-branes to the $IR$ region in the near-horizon geometry. The fundamental matter resides in the vector multiplet of ${\mathcal N}=4$,
  it is in the fundamental/antifundamental representation of the $SU(N)$ group and in the anti-fundamental/fundamental representation of the $SU(M)$ flavor group. The mass of
  the quarks is proportional to the separation between the two stacks of $D$-branes. Note that when we take the separation to zero the gauge group
  becomes $SU(N+M)$, and the fundamental matter of $SU(N)$ becomes part of the adjoint of the $SU(N+M)$ group. Thus, in this limit massless
  quarks have by construction essentially the same properties as gluons.

\begin{figure}[htbp]
\begin{center}
\epsfig{file=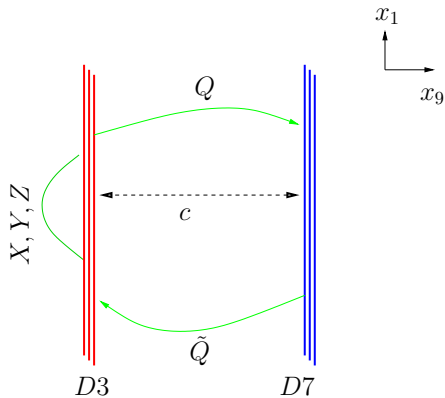,scale=0.7}
\caption{The $D3-D7$ system.} \label{D3D7}
\end{center}
\end{figure}

  The second and more interesting way to add fundamental matter is by
  adding $M$ $D7$-branes  to $N$ $D3$-branes \cite{Karch:2002sh}.\footnote{See also \cite{Kruczenski:2003be} for more details.} Again, we take $N$ large but $M$ finite. SUSY is broken to ${\mathcal N}=2$. The
  fundamental matter comes from strings stretching between the $D7$-branes and the $IR$ region of the near-horizon geometry. These fields
  sit in the hypermultiplet of ${\mathcal N}=2$ and transform in the fundamental/anti-fundamental representation of the $SU(N)$ group and in the
   anti-fundamental/fundamental    representation of the $SU(M)$ flavor group. The mass of the quarks is given by the separation of
    the $D7$-branes from the $D3$-branes (see figure \ref{D3D7}). In the
  near horizon limit the $D7$ branes fill the $AdS_5$ part of the space up to a specific $IR$ cutoff given by the mass of the quarks. We will be
  interested in massless quarks in what follows.

\begin{figure}[htbp]
 \begin{center}
 $\begin{array}{c@{\hspace{1.6in}}c}
\epsfig{file=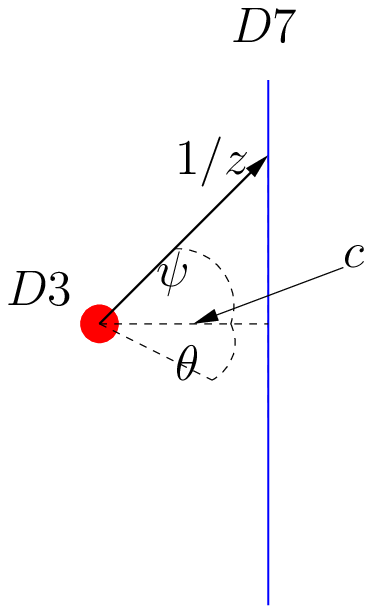,scale=0.7} & \epsfig{file=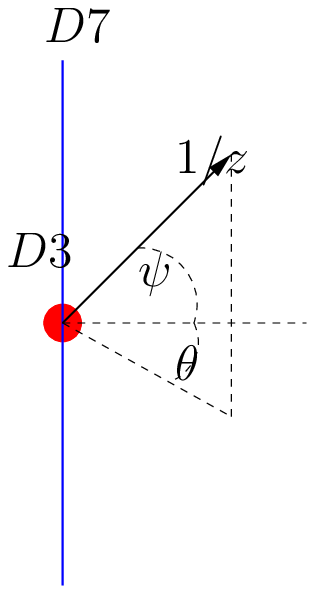,scale=0.7}
\\ [0.2cm]
(a)&(b)\\ [0.2cm]
\end{array}$
\caption{The $D3-D7$ system before backreaction considered. (a) The massive case, here the $D7$ wraps
the sphere in a non-trivial way, $\cos(\psi) = c\cdot z$. (b) The massless case, the $D7$ brane
lies along the $\psi=\half \pi$ direction, and spans all values of $z$.  } \label{D3D7bb}
\end{center}
\end{figure}

  Let us specify some details of this theory. We take the stack of $M$ $D3$-branes to span the directions $x_{0,..,3}$. The stack of $M$ $D7$-branes
  is sitting at $x_8=0$ and $x_9=c$ and spans the remaining space-time directions. Note that the mass of the quarks
  is proportional to the parameter $\frac{c}{l_s^2}$. Further, we define
  \be\label{r}
  \frac{1}{z^2}\equiv \sum_{i=4}^9x_i^2,
  \ee
  which becomes the radial direction of the $AdS_5$ factor and the sub-manifolds of constant $z$ span the $S^5$ part of the
  the near horizon geometry. The metrics on the $AdS_5$ and the $S^5$ factors are given by,
  \be\label{ads_metric}
  ds_{AdS}^2=\frac{R^2}{z^2}\biggl(dz^2+\sum_{i,j=0}^3\eta^{ij}dx_idx_j\biggr),\quad
  ds_{S^5}^2=R^2\biggl(d\psi^2+\cos^2(\psi)d\theta^2+\sin^2(\psi)d\Omega_3^2\biggr),
  \ee
  where the $D7$-branes wrap the $S^3$ factor as \be\label{wrap}\cos(\psi)=c\cdot z.\ee
The geometrical setup giving rise to (\ref{wrap}) is depicted in figure \ref{D3D7bb}(a)

   Note that for non zero $c$ the $D7$-branes
  fill the $AdS_5$ space for $z\leq 1/c$, and only for $c=0$ the whole $AdS$ space is covered (this special case is depicted in figure \ref{D3D7bb}(b)).
  We will specify now to the massless case by taking $c=0$. The superpotential in this case is given by
  \be\label{superpot}
  W=X\left[Y,Z\right]+YQ\tilde Q,
  \ee
  where $Q$ and $\tilde Q$ form together a hypermultiplets of ${\mathcal N}=2$ coming from strings stretching
  between the $D7$ and the $D3$ branes. Note that the fundamental matter couples to a single ${\mathcal N}=1$ adjoint scalar $Y$.
  The choice of this scalar is dictated by the choice of the $S^3$ inside the $S^5$ which the $D7$-branes wrap, given in our case by \eqref{wrap}.

\section{$n$-point quark scattering}\label{sect_Tdual}
In this section we explain the minimal area problem which is relevant for computing quark amplitudes at strong
coupling. We introduce quarks via the $D3-D7$ system discussed above. The massless quarks come from strings
connecting a probe $D3$ brane located at $z=\infty$ and the probe flavor $D7$ branes. The $D7$ branes span the
whole $AdS$ factor and are localized at $\psi=\half \pi$ on the sphere. We regulate the possible $IR$
divergences with dimensional regularization. The dimensionally continued metric of $p=3-2\epsilon$ branes is
given by (the $r$ coordinate is given by $r=\frac{R^2}{z}$) \cite{Alday:2007hr} \be\label{dimreg}
ds^2&=&\frac{r^2 dx_D^2}{\sqrt{\lambda} \left[2\pi\frac{\mu}{r} e^{\tilde \gamma}\right]^{\e}
\sqrt{\Gamma(2+\e)}}+\sqrt{\lambda} \left[2\pi\frac{\mu}{r} e^{\tilde \gamma}\right]^{\e}
\sqrt{\Gamma(2+\e)}\frac{dr^2}{r^2}+\sqrt{\lambda} \left[2\pi\frac{\mu}{r}  e^{\tilde
\gamma}\right]^{\e} \sqrt{\Gamma(2+\e)}d\Omega^2_{9-D},\nonumber\\
&&\tilde\gamma=-\half \Gamma'(1),\qquad \lambda=g^2N,\qquad
D=4-2\e. \ee The relevant sign of the regulator is $\e<0$ (the
reason is that IR divergences are regulated by increasing the
dimension of space time), $\lambda$ is the dimensionless $4d$ 't
Hooft coupling, and $\mu$ is the $IR$ scale of dimensional
regularization.

In principal, one has to specify the position of the $D3$ brane on
the $S^5$ since there is an invariant angle between the $D3$ and
the $D7$. Irrespectively of this angle, because of the infinite
rescaling of the energy
\be E_{\mathcal N =4}=\frac Rz E_{10 D}, \ee
the quarks in the gauge theory are massless as long as the $D3$
sits at $z=\infty$. Thus, the (dimensionally regularized)
scattering process we perform in field theory is independent of
this angle, and so should be the final result we get from string
theory. For simplicity we may put our $D3$ brane at
$\psi=\frac12\pi$, on top of the $D7$ branes.

We wish to emphasize that this angle independence is necessary for
the consistency of the prescription \footnote{We are grateful to
O. Aharony for invaluable discussions on this.} not only in the
$D3-D7$ system but also in the case of gluon scattering. This
issues was not relevant in \cite{Alday:2007hr}, but it certainly
becomes relevant if one takes out more than a single $D3$ out of
the stack. One may decide to scatter massless gluons connecting
the two $D3$ branes which were removed from the stack. There must
not be any angle dependence due to color decomposition in the
field theory process.\footnote{It will be interesting to
understand better how this comes about.}

The main a priori difference between scattering of gluons and scattering of quarks is that the $D7$ branes are
not localized along the $z$ direction. This implies that the boundary of the disc on which the string vertex
operators are inserted may not be restricted to $z=\infty$; it is a Neumann direction.

The  power and the simplicity of the Alday-Maldacena mechanism is
the translation of the scattering problem into a minimal area
problem with a prescribed boundary. The translation between these problems
goes through a formal T-duality procedure of some of the $AdS_5$
coordinates (all except the radial one). The T-dual coordinates are defined to be
\be\label{tdual} \d_\a y^\mu=i\frac{r^2}{R^2}\e_{\a\beta} \d^\beta
x^\mu. \ee The T-dual space has still an $AdS_5$
metric
 which has the following dimensionally regularized form
\be\label{Tdualmetric} ds^2&=&\sqrt{\lambda}  \left[2\pi\mu
e^{\tilde \gamma}\right]^{\e}
\sqrt{\Gamma(2+\e)}\frac{dr^2+dy_D^2}{r^{2+\e}}+\sqrt{\lambda}
\left[2\pi\mu e^{\tilde \gamma}\right]^{\e}
\sqrt{\Gamma(2+\e)}\frac{d\Omega^2_{9-D}}{r^\e}. \ee After this
T-duality the scattering problem translates to a minimal area
problem with the boundary of the surface fixed by straight
light-like segments.

We are looking for a classical solution of the world-sheet
embedding into $AdS$ space around an insertion of a quark vertex
operator. Consider the following vertex operator
\be\label{vertex} V\sim
e^{ik^\mu x_\mu}, \qquad k^2=0, \ee
where we have truncated to the lowest lying excitation in the $x$ fields (in other words we have not included any function of the derivatives of $x$). The exponent is determined by shift symmetries in $x$ and the
$AdS$ metric in the Poincare patch is given by \eqref{ads_metric}.
 Thus, the classical action of the string with such a vertex
operator
 is given by
\be\label{class_string} S=\int
d^2w\left[\frac{R^2}{2\pi\alpha'z^2}\left(\d z\bar\d z+\d
x^\mu\bar\d x_\mu\right)+ik^\mu x_\mu\delta^{(2)}(w,\bar
w)\right], \ee and the equations of motion are given by
\be\label{class_eom} \bar \d\d x^\mu -\frac{1}{z}(\d z\bar\d
x^\mu+\bar \d z \d x^\mu)=i\frac{\pi\alpha' z^2}{R^2}k_\mu
\delta^{(2)}(w,\bar w),\qquad \bar \d\d z+\frac{1}{z}\left(\d
x_\mu\bar\d x^\mu-\d z\bar\d z\right)=0. \ee Indeed, a solution to
this coupled system can be easily found in the massless case
\be\label{solution} z=z_0 \hspace{2em} x_\mu \sim
i\frac{\alpha'z_0^2}{2R^2}k_\mu \ln|w|^2. \ee It can be easily
seen to satisfy the classical constraint $L_0=0$. Of course, up to
now there is absolutely no technical difference between the
Dirichlet-Dirichlet case considered in \cite{Alday:2007hr} and the
Dirichlet-Neumann case we consider here. The solution
(\ref{solution}) satisfies both types of boundary conditions, and
describes a massless string which does not stretch in the z
direction at all.

\begin{figure}[htbp]
\begin{center}
\epsfig{file=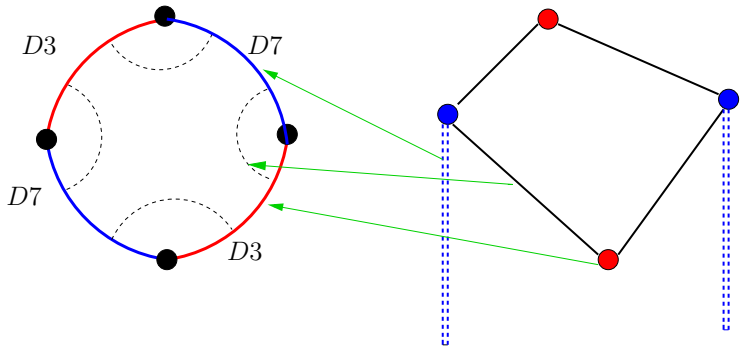,scale=0.7} \caption{The disc four point
amplitude before (left) and after (right) the action of the
$T$-duality. The spike singularities which arise if $z$ is allowed to vary along the blue lines on the left hand side figure, are depicted as blue dashed
lines.} \label{TDuality}
\end{center}
\end{figure}

A difference does arise if we recall that the full scattering problem has many vertex operators, and the embedding along the boundaries connecting different vertex operators can be non trivial. In the case of  \cite{Alday:2007hr} this is excluded automatically by the Dirichlet-Dirichlet boundary conditions, but not in our case. Remembering that near each vertex operator the consistent solution has constant $z$, a variation of $z$ between different vertex operators results in a singular surface describing the boundary condition in the T dual frame. This situation is depicted in figure \ref{TDuality}. The reason for these spike singularities in the boundary surface is that the T dual $y$'s are constant along the boundary line connecting different vertex operators while the $z$'s vary by assumption. The opposite situation occurs in the neighborhood of each vertex operator.

 Thus, in the end, the calculation of the
quark scattering at strong coupling necessarily amounts to a
solution of a minimal area problem with fixed boundary formed by
light-like segments, perhaps with a spike. After this paper has
appeared several arguments were presented in
\cite{McGreevy:2007kt} in favor of the solution with the
spikes.\footnote{ In a previous version of this paper we assumed
that there should be no spikes. However, in light of the arguments
presented in \cite{McGreevy:2007kt} the spikes should be taken
into consideration.}

\section{ ${\mathcal N}=2$ conformal field theory }\label{N=2conf}
In this section we analyze a closely related case, the ${\mathcal
N}=2$ conformal field theory of
\cite{Banks:1996nj,Aharony:1996en,Douglas:1996js}.\footnote{We are
grateful to S.Yankielowicz who brought this model to our
attention.}
 It consists of
$USp(2N)$ gauge theory with one hypermultiplet in the
antisymmetric representation and four hypermultiplets in
the fundamental representation. It can be realized by probing F theory
\cite{Vafa:1996xn} compactified on K3 at a special point in moduli
space with $D3$ branes. At this special point, the K3 is actually the orbifold
$\mathbb{T}^4/\mathbb{Z}_2$ \cite{Sen:1996vd}. In ten dimensions, in the vicinity of one of the four fixed points,
this corresponds to a $\mathbb{Z}_2$ orientfold of IIB string
theory in flat space which consists of an $O7^-$ plane and four
$D7$ branes on top of it. This cancels the tadpole locally,
allowing the string coupling to be arbitrary. $D3$ branes probing
this configuration carry the ${\mathcal N}=2$ theory described
above where the fundamental hypermultiplets arise from D3-D7
strings. In the origin of the moduli space of D3 branes all the
fields are massless and the theory is exactly conformal.

In the large N limit with large 't-Hooft coupling this is described by the dual IIB string theory background \cite{Aharony:1998xz}

\be\label{ads-N=2_metric}
  ds^2=\frac{R^2}{z^2}\biggl(dz^2+\sum_{i,j=0}^3\eta^{ij}dx_idx_j\biggr)+R^2\biggl(d\psi^2+\cos^2(\psi)d\theta^2+\sin^2(\psi)d\Omega_3^2\biggr),
  \ee
where the essential difference from the usual case is that $\theta$ has periodicity $\pi$ instead of $2\pi$ and that
the doublet of two forms $(B^{(2)},C^{(2)})$ goes to $(-B^{(2)},-C^{(2)})$ upon $\pi$ rotation in $\theta$.
For this specific singularity, the axion-dilaton field doesn't have non trivial monodromies since the $SL(2,\mathbb{Z})$
element defining the fiber in F theory is actually an identity in $PSL(2,\mathbb{Z})$.

We would like to emphasize that the presence of dynamical quarks implies that the solution (\ref{ads-N=2_metric}) actually contains in it
$D7$ branes whose back-reaction was taken into account. In a sense, the background (\ref{ads-N=2_metric}) should be thought of as already having $D7$ branes, and the angle $\theta$ should be thought of as the angular variable in the transverse space to the
$D7$s which are localized in $\psi=\pi/2$, wrap the remaining $S^3$ and are extended in all the directions of $AdS_5$ (the situation is practically the same as in figure \ref{D3D7bb} only that one has to add an orientfold plane and the branes come in pairs).

For our purpose it remains to emphasize again that this theory
contains dynamical fields in the fundamental representation, and
that both what we have done above and the gluon scattering of
\cite{Alday:2007hr} can be repeated {\it verbatim}. In particular,
color decomposition holds in this case as well (with a small
technical difference due to the fact that the fundamental
representation is equivalent to its conjugate) and one can take a
D3 (and its mirror which is reached after a rotation in $\theta$)
out of the stack , replacing the stack by the gravity background
(\ref{ads-N=2_metric}). Scattering of gauge bosons in this system
corresponds to scattering in the $U(1)$ of the coulomb branch of
$USp(2N)$
\begin{equation}\label{coulomb}USp(2N-2)\times U(1)\hookrightarrow USp(2N).\end{equation}
This branch may be reached by turning VEVs to the scalars in the vector multiplet or by turning VEVs of scalars in the antisymmetric hypermultiplet.
The difference is that the latter moves the branes parallel to the $D7$ branes while the former moves them in
transverse directions.\footnote{ Note that if there is no movement at all in transverse directions to the $D7$, then other symmetry breaking patterns become possible, for instance $USp(2N-2)\times USp(2)$. This subtlety will not change anything in our considerations.}
Since gluons in the $U(1)$ are just Dirichlet-Dirichlet strings of the $D3$ brane, it does not matter where it sits on the orientfolded $S^5$ (as in the oriented case, there must be "angle independence" here as well) and
the T duality of \cite{Alday:2007hr} is performed in exactly the same way, with the same result for the reduced gauge invariant amplitude of gluon scattering.

Scattering of fundamental hypermultiplets is very natural in this system, since no $D7$ branes have to be
introduced by hand. On the coulomb branch (\ref{coulomb}) the hypermultiplets are massive and are described by
Dirichlet-Neumann strings with an end on the $U(1)$ $D3$ brane and the other end at any point in
$AdS$.\footnote{Of course, these strings must also stretch in the transverse space to $AdS$ if we reached the
coulomb branch with vector multiplet expectation values.} To get the theory in the origin of the moduli space we
bring the $D3$ brane back to the origin. It is clear that the minimal area problem we get here is the same as
the one  we got for the $\mathcal{N}=2$ theory analyzed in the previous section, implying a universal result at
strong coupling.
\section*{Acknowledgments}
We would like to thank I.~Adam, O.~Bergman, S.~Itzhaki,
M.~Lippert, Y.~Oz, D.~Reichmann, J. Sonnenschein and in particular
O.~Aharony, M.~Berkooz, A.~Schwimmer, S.~Yankielowicz
 for many illuminating,
useful and interesting discussions as well as for commenting on
the manuscript. The work of Z.K. was supported in part by the
Israel-U.S. Binational Science Foundation, by the Israel Science
Foundation (grant number 1399/04), by the European network
HPRN-CT-2000-00122, by a grant from G.I.F., the German-Israeli
Foundation for Scientific Research and Development, by a grant
of DIP (H.52), and by the Minerva Center for Theoretical Physics.  The work of S.S.R. is supported in part by Israel
Science Foundation under grant no 568/05.

\end{document}